# Coherent manipulation of magnetization precession in ferromagnetic semiconductor (Ga,Mn)As with successive optical pumping

Y. Hashimoto and H. Munekata

*Imaging Science and Engineering Laboratory, Tokyo Institute of Technology,*

*4259-R2-57 Nagatsuta, Midori-ku, Yokohama 226-8502, Japan*

(e-mail of corresponding author : hiro@isl.titech.ac.jp)

**Abstract**

We report dynamic control of magnetization precession by light alone. A ferromagnetic (Ga,Mn)As epilayer was used for experiments. Amplitude of precession was modulated to a large extent by tuning the time interval between two successive optical pump pulses which induced torques on magnetization through a non-thermal process. Nonlinear effect in precession motion was also discussed.

It was found in 2005 that optical excitation of ferromagnetic semiconductor (Ga,Mn)As with 100-fs laser pulses induces precession of magnetization that last up to around 1 ns [1]. Since then, the photo-induced precession has been studied to develop knowledge on spin dynamics of carrier-mediated ferromagnetism [2-5]. It has been reported that a torque that works on ferromagnetically-coupled Mn spins is produced by a slight tilt of a *p-d*-exchange-induced magnetic anisotropy field with photo-generated extra holes of the order of $10^{16}$ - $10^{17}$ cm$^{-3}$ [5]. This fact suggests that the motion of magnetization may be controlled precisely by the sequence of optical pulses. It may lead us to the opportunity of studying an effect of memory in optical signal processing [6].

Coherent control of precessions of individual and ordered spins has already been reported for a paramagnetic, II-VI diluted magnetic semiconductor [7] and an antiferromagnetic insulator [8], respectively. However, coherent control of precession of magnetization has not been demonstrated yet. In this letter, we show, with both experimental data and model calculation, that the amplitude of magnetization precession in a ferromagnetic semiconductor (Ga,Mn)As can either be enhanced or suppressed by tuning the interval between two successive pump pulses.

Experiments were carried out by measuring time-resolved magneto-optical (MO) signals with pump and probe technique. As shown schematically in Fig. 1(a), a beam splitter and the second delay stage were added to the previous experimental setup [5] to yield the first and the second, 76-MHz pump trains of equal intensity (3.4 μJ/cm$^2$/pulse). Intensity of a probe pulse train was 0.34 μJ/cm$^2$/pulse. All three pulse trains, whose polarization plane was aligned along the crystallographic [010] axis, were focused by a single lens to a spot of 150-μm diameter on the surface of a 100-nm thick $Ga_{0.98}Mn_{0.02}As$ sample. Spatial overlap of pulses at the sample surface was optimized by monitoring light intensity through a pinhole of 50 μm diameter. The sample was first magnetized along the crystallographic [100] (in-plane) at 10 K, after which measurements were carried out at the same temperature without an external field. Further detailed experimental procedure has been described in ref.5.

Two temporal MO profiles obtained by the excitation with the first pump pulse (solid line) and the second pump pulse (dashed line) are shown in Fig.1(b). Both profiles show qualitatively similar behavior, consisting of relatively large initial amplitude due to lateral motions of magnetization (*M*) followed by the natural damping through which vertical *M* component is yielded [5]. Rigorously stated, these two profiles are not quantitatively identical, as evidenced by the differences in oscillation frequency ($\Delta\omega$ = 0.28 GHz) and amplitude. The difference, which was attributed to a slight change in spatial overlap of pump-probe pulses between the two experiments, could not be solved completely within the limit of our best effort with the pinhole procedure. We found, however, that the observed degree of difference between the first- and the second-pump excitations did not cause serious difficulties in analyzing experimental data obtained by the two-pumps excitation. Influence of magnetic inhomogeneity (e.g., multi-domains) in a 150-μm spot was negligibly small, since changing the sampling position did not vary the data shown in Fig.1(b).

Figure 2 shows MO signals observed by the excitation with two successive pump pulses having the phase interval of $\Omega \cdot t_{int}$ = 0, π, and 2π.  Here, $\Omega$ is the frequency of precession induced by the first-pump excitation and $t_{int}$ is the time interval between the two pump pulses.   One finds that oscillation in MO signals is either suppressed or enhanced depending on the $\Omega \cdot t_{int}$ value.   For $\Omega \cdot t_{int}$ = 0, for which two pumps excite the sample simultaneously, a temporal MO profile exhibits similar oscillatory behavior with twice as large amplitude as those shown in Fig.1(b).   For $\Omega \cdot t_{int}$ = π ($t_{int}$ = 100 ps), the oscillation is suppressed significantly after excitation with the second pump pulse, whereas a cliff-like profile remains in the MO profile.   As will be discussed in the later paragraph, the observed decay is attributed to a recovery of $M$ toward its equilibrium position.   For $\Omega \cdot t_{int}$ = 2π ($t_{int}$ = 200 ps), the oscillation amplitude is strongly enhanced after excitation with the second pump, as manifested itself in signals at around 350 ps. Experimental results indicate that the photo-induced precession of $M$ is successfully manipulated by means of the second-pump pulse.

A model which successfully reproduced magnetization precession with one-pulse excitation has been given by the following four equations [5]:

$$\frac{d\vec{M}(t)}{dt} = -\gamma[\vec{M}(t) \times H_{eff}(t)] + \frac{\alpha}{M_s}[\vec{M}(t) \times \frac{d\vec{M}(t)}{dt}] \qquad (1)$$

$$\begin{aligned}H_{eff}(t) &= (H_0 \cos\theta(t), 0, H_0 \sin\theta(t)) \\ &\approx (H_0, 0, H_0\theta(t))\end{aligned} \qquad (2)$$

, for which $\theta(t) \ll \pi$.

$$\theta(t) = \theta_0(1-\exp(-t/50ps))\exp(-t/500ps) \qquad (3)$$

and

$$\Theta_{th}(t) = \frac{-M_y(t) + 0.45 M_z(t)}{M_s} [radian] \qquad (4)$$

Essential point of this model is that a change in an effective magnetic anisotropy field $H_{eff}$ induced by the optical excitation is expressed in terms of a temporal tilt of $H_{eff}$ as represented by eqs. (2) and (3), and magnetization $M$ starts its motion on the basis of Landau-Lifshitz-Gilbert (LLG) equation (1). For the un-annealed, $x = 0.02$ sample, the adjusting parameter $\theta_0 = 10$ mrad., the damping factor $\alpha = 0.16$, and the anisotropy field $\mu_0 \cdot H_{eff} = 0.21$ Tesla [5]. The motion of $M$ thus calculated was then converted into MO signals $\Theta_{th}(t)$, the rotation angle of the polarization plane of the reflected probe pulse, as expressed by eq. (4). Here, the first term represents in-pane ($y$) component through magnetic linear dichroism, and the second term out-of-plane ($z$) component through polar Kerr rotation. A change in the in-plane, $x$ component (parallel to the initial $M$) is two orders of magnitude smaller than $y$ and $z$ components, and ignored here.

A model for the two-pump excitation is generated by adding the influence of the second pump which arrives at the sample surface with the time delay $t_{int}$ with respect to the first pump. The photo-induced directional change of an anisotropy field is given by $H_{eff}(\theta_{1,2})$ in which $\theta_{1,2}$ is a linear combination of photo-induced change due to the first- and the second-pump pulse, $\theta_1$ and $\theta_2$, respectively.

$$\theta_{1,2}(t, t_{int}) = \theta_1(t) + \theta_2(t, t_{int}) \qquad (5)$$

Shown in Figs.3 (a), (b), and (c) are the calculated motions of $M(t)$ with two-pump excitation in the $M_y$ vs. $M_z$ space for three different phase intervals of $\Omega \cdot t_{int} = 0, \pi,$ and $2\pi$, respectively. Temporal profiles of $\theta_{1,2}$ for each $\Omega \cdot t_{int}$ are shown in Fig.3 (d). Notice that the amplitude of precession is altered significantly by the value of $\Omega \cdot t_{int}$; the oscillation amplitude becomes twice as large for both $y$ and $z$ directions as that of one-pump excitation for $\Omega \cdot t_{int} = 0$; whereas, for $\Omega \cdot t_{int} = \pi$, oscillatory motion is suppressed after the second pump, especially for $y$ component. In this case, a change in $z$ component dominates the motion of $M$. As for $\Omega \cdot t_{int} = 2\pi$, the oscillatory motion is enhanced after the second pump. Temporal MO profiles based on the motions of $M(t)$ are produced by using eq.(4), and are plotted in Fig.2 with dots. Experimental data (solid lines) are well reproduced by the model calculation within the limit of slight but noticeable difference in the pump-and-probe experimental condition between the first and the second pumps.

Frequency modulation may be possible in view of nonlinear nature of the LLG equation. With the nonlinear contribution, MO signals $\Theta_{1,2}$ is expressed as:

$$\Theta_{1,2}(t, t_{int}) = \Theta_1(t) + \Theta_2(t, t_{int}) + \Theta_{1,2}(t, t_{int}) \tag{6}$$

Here, $\Theta_1(t)$, $\Theta_2(t, t_{int})$, and $\Theta_{1,2}(t, t_{int})$ are the experimental MO signals induced by the first- and the second-pumps, and the nonlinear term, respectively. $\Theta_{1,2}(t, t_{int})$ should be detected with special care, since, rigorously stated, the sample temperature at the quasi steady state during the experiment is different between one-pump and two-pumps experiments, although the difference is small ($\Delta T \sim 1$ K for one pump experiment [5]). To circumvent this problem, only the first-pump pulse was optically chopped (378 Hz), and MO signals locked in this frequency were detected during the two-pumps experiment. The signals are expressed as:

$$\Theta_{1, lock}(t, t_{int}) = \Theta_{1,2}(t, t_{int}) - \Theta_2(t, t_{int}) = \Theta_1(t) + \Theta_{1,2}(t, t_{int}) \tag{7}$$

Here, the right-hand-side of eq.(7) results from eq.(6). If experimentally observed $\Theta_{1,\text{lock}}(t, t_{\text{int}})$ depends on the value of $t_{\text{int}}$, it must be caused by the $\Theta_{1,2}(t, t_{\text{int}})$ term. Figure 4 shows four $\Theta_{1,\text{lock}}(t, t_{\text{int}})$ profiles observed with different $t_{\text{int}}$ values, which exhibits that those profiles are independent of $t_{\text{int}}$. This fact indicates that $\Theta_{1,2}(t, t_{\text{int}})$ is negligibly small under the present weak excitation condition. Indeed, model calculation has showed that the effect of nonlinearity is too small to be detected within the limit of the present experimental setups. About two orders of magnitude larger MO signals would be needed to detect the the $\Theta_{1,2}(t, t_{\text{int}})$ term. Experiments with higher pump power and/or with samples having larger MO coefficients would be desired.

In summary, amplitude modulation of precession motion of magnetization in (Ga,Mn)As has been achieved by controlling the time interval between the two successive optical pump pulses. The observed behaviors have well been explained in terms of the linear combination of photo-induced torque caused by each pump pulse. Results have showed that technique of ultrafast, weak optical excitation can be applied to coherently control precession motion of magnetization, at least in ferromagnetic semiconductors.

This work is supported in part by Grant-in-Aid for Scientific Research from JSPS (No. 17206002), MEXT (No.19048020), and by the NSF-IT program (DMR-0325474).

**References**

[1] A. Oiwa, H. Takechi, and H. Munekata, J. Supercond. Nov. Magn. **18**, 9 (2005).

[2] D. M. Wang, Y. H. Ren, X. Liu, J. K. Furdyna, M. Grimsditch, and R. Merlin, Phys. Rev. B **75**, 233308 (2007).

[3] J. Qi, Y. Xu, N. H. Tolk, X. Liu, J. K. Furdyna, and I. E. Perakis, Appl. Phys. Lett. **91**, 112506 (2007).

[4] E. Rozkotová, P. Němec, P. Horodyská, D. Sprinzl, F. Trojánek, P. Malý, V. Novák, K. Olejník, M. Cukr, and T. Jungwirth, Appl. Phys. Lett. **92**, 122507 (2008).

[5] Y. Hashimoto, S. Kobayashi, and H. Munekata, Phys. Rev. Lett. **100**, 067202 (2008).

[6] D.K. Huneter, M.C. Chia, and I. Audonovic, J. Lightwave Technol. **16**, 2081 (1998).

[7] R. Akimoto, K. Ando, F. Sasaki, S. Kobayashi, and T. Tani, Phys. Rev. B **56**, 9726 (1997).

[8] A. V. Kimel, A. Kirilyuk, F. Hansteen, R.V. Pisarev, and T. Raising, J. Phys.: Condens. Matter **19**, 043201 (2007).

**Figure captions**

Figure 1:

(color online) (a) Schematic illustration of experimental configuration, and (b) temporal MO profiles caused by the first-pump pulse (solid line) and the second-pump pulse (dashed line).

Figure 2:

(color online) Temporal MO profiles obtained by two pump experiments with three different pump intervals $\Omega \cdot t_{int}$ = 0, $\pi$, and $2\pi$ (plots with dots). Arrows depict the time of excitation with the second-pump pulse. Temporal profiles obtained by model calculation are shown by solid lines.

Figure 3

(color online) Precessional motions of magnetization in $M_y$ - $M_z$ parameter space calculated for (a) $\Omega \cdot t_{int}$ = 0, (b) $\pi$, and (c) $2\pi$. Arrows depict the time of excitation with the second-pump pulse. (d) Temporal profiles of $\theta_{1,2}$ for three different $\Omega \cdot t_{int}$ of 0 (solid line), $\pi$ (dashed line), and $2\pi$ (dotted line).

Figure 4

(color online) Temporal MO profiles of $\Theta_{1, lock}(t, t_{int})$ at $\Omega \cdot t_{int}$ = $-\pi$ (circles), 0 (triangles), $\pi$ (squares), and $2\pi$ (diamonds). Arrows represent the time of excitation with the second-pump pulse.

**Figures**

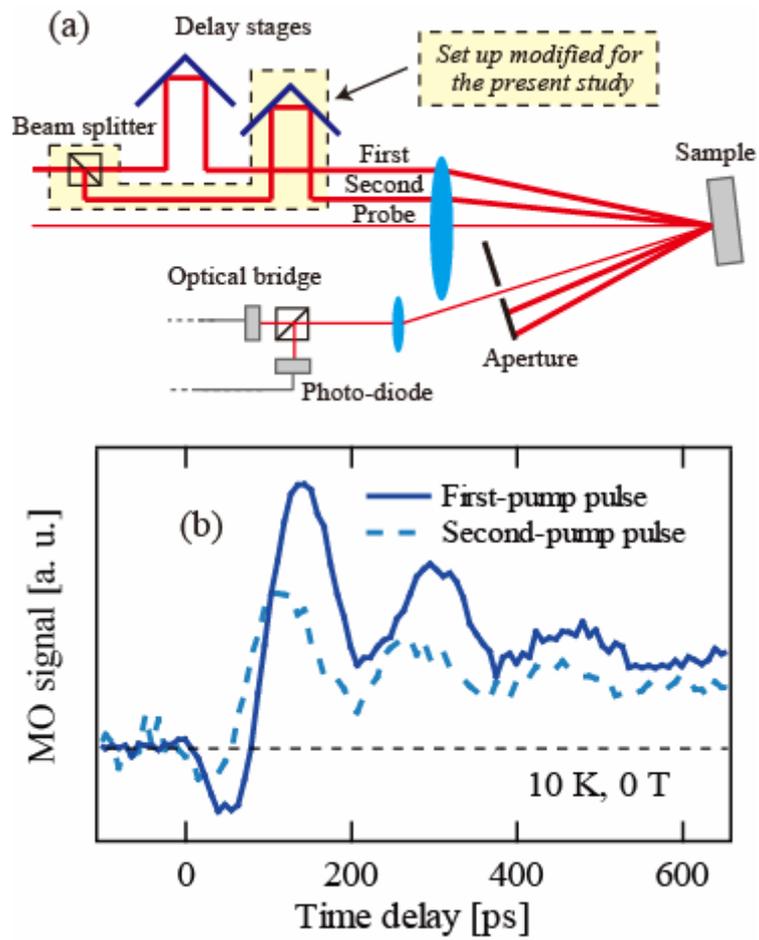

Figure 1    Y. Hashimoto and H. Munekata

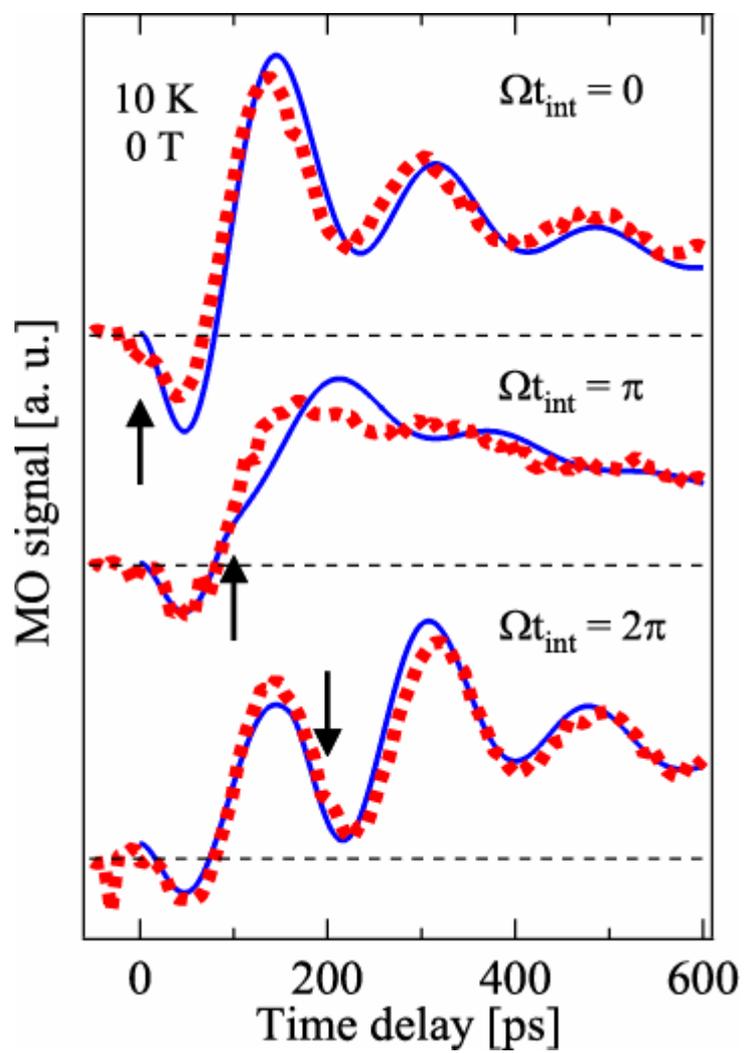

Figure 2    Y. Hashimoto and H. Munekata

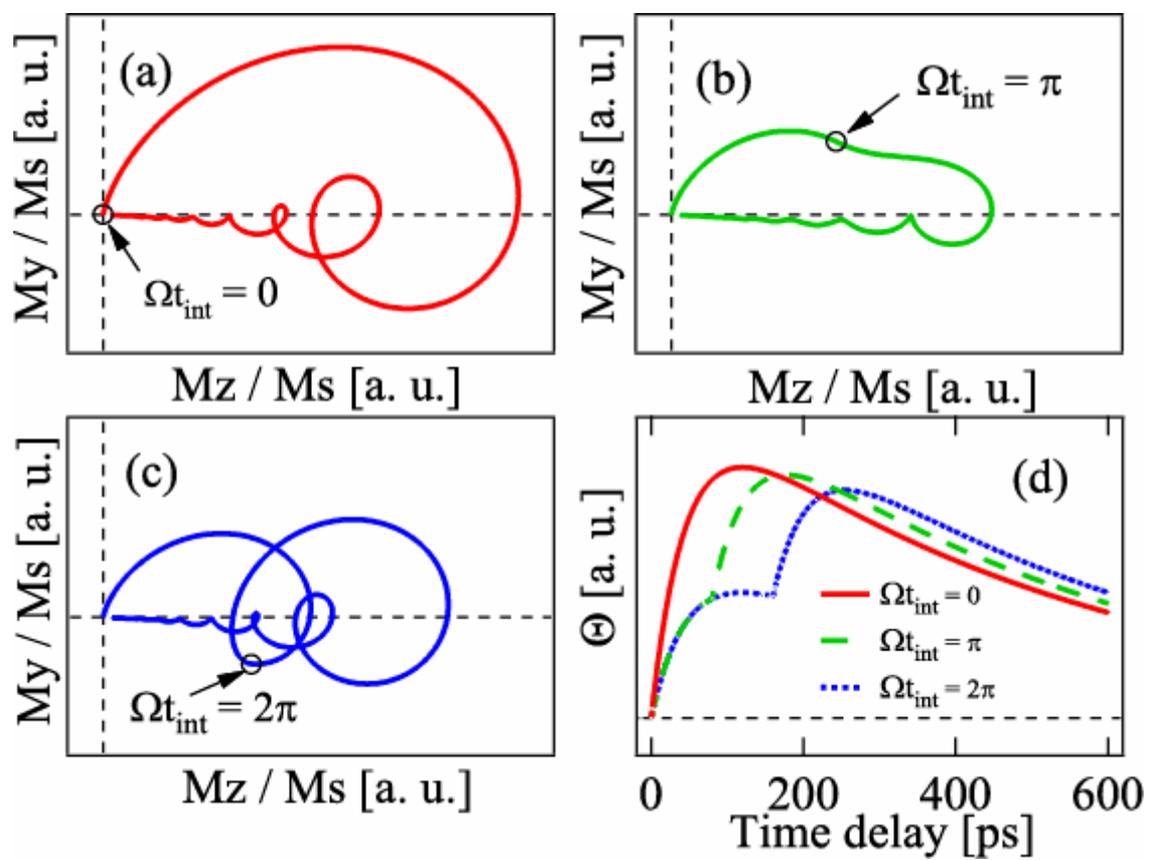

Figure 3    Y. Hashimoto and H. Munekata

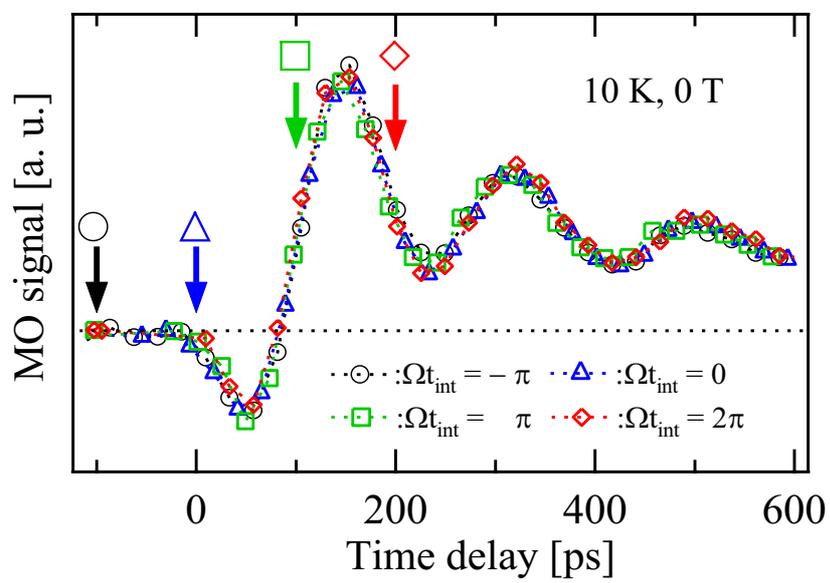

Figure 4    Y. Hashimoto and H. Munekata